\documentclass[aps,prb,twocolumn,nofootinbib]{revtex4}
\usepackage{amsmath} 
\usepackage{amsfonts} 
\usepackage{graphicx} 
\usepackage{stix}
\usepackage[export]{adjustbox}

\begin{document}
\title{Simulating spin dynamics with quantum computers}
\author{Jarrett L. Lancaster}
\email[Electronic mail: ]{jarrett.lancaster@mma.edu}
\affiliation{Department of Physics, High Point University, One University Parkway, High Point, NC 27262} 
\altaffiliation{Currently at Department of Arts \& Sciences, Maine Maritime Academy, One Pleasant Street, Castine, ME 04420} 
\author{D. Brysen Allen}
\affiliation{Department of Physics, High Point University, One University Parkway, High Point, NC 27262} 
\date{\today}

\begin{abstract}
IBM quantum computers are used to simulate the dynamics of small systems of interacting quantum spins. For time-independent systems with fewer than three spins, we compute the exact time evolution at arbitrary times and measure spin expectation values and energy. It is demonstrated that even in such small systems, one can observe the connection between conservation laws and symmetries in the model. Larger systems require approximating the time-evolution operator, and we investigate the case of $N=3$ spins explicitly. While it is shown to be unfeasible to use such devices to probe such larger systems without more advanced algorithms or reliable error correction, we demonstrate that the quantum circuit simulator is an easy-to-use method for studying spin dynamics in systems with $N\sim\mathcal{O}(10)$) spins. The computations presented provide an interesting experimental component to the standard treatment of quantum spin in an undergraduate quantum mechanics course.
\end{abstract}

\maketitle

\section{Introduction}\footnote{Accepted by {\it Am. J. Phys.}. After it is published, it will be found at:\\ \href{https://pubs.aip.org/aapt/ajp}{https://pubs.aip.org/aapt/ajp}}
The field of quantum computation has experienced tremendous growth in recent years. In 1982, Richard Feynman suggested that (then-hypothetical) quantum computers were the only practical way to simulate quantum mechanical systems with many degrees of freedom.\cite{Feynman}  Over the next several decades, quantum computing algorithms were developed that may play a potentially valuable role in an efficient factorization scheme.~\cite{Shor} Such algorithms present a threat to current cryptographic systems because the security of the widely-used Rivest–Shamir–Adleman (RSA)~\cite{RSA} cryptographic scheme rests upon the computational difficulty of factoring a product of two large, prime numbers. So-called noisy, intermediate-scale quantum (NISQ) devices~\cite{Preskill} are widely available from providers such as Rigetti,\cite{Rigetti} IonQ,\cite{IonQ} and IBM,\cite{IBMQ} though the currently-available devices are still far from capable of these advanced types of computations. The question for the foreseeable future appears to be: what kinds of problems are these NISQ devices capable of solving?

While many potential uses for NISQ-era hardware have been proposed,\cite{Alexeev} one under-appreciated use for these devices is as a learning tool for quantum mechanics. Presently available quantum computers are ideally suited for exploring spin physics in small systems,~\cite{Brody} such as those that would be explored theoretically in a typical undergraduate course on quantum mechanics. Recent papers in this journal have showcased a variety of quantum information applications suitable for undergraduate students. Examples include exploring basic problems future quantum computers can conceivably be used to solve,\cite{Candela} employing the Grover search algorithm on actual quantum hardware,\cite{Kain} and applying basic error mitigation to improve results on currently available hardware.\cite{Johnstun} A particularly interesting physical application of this technology used quantum computers to measure spin correlations.\cite{Brody} Given the free availability actual quantum hardware from IBM, the simulation of spin systems provides an incredible opportunity to students. In other words, there is now a no-cost experimental component to complement the fairly abstract theory one finds in a typical, undergraduate course in quantum mechanics.

The main purpose of this paper is to describe several types of spin simulations involving the time evolution of arbitrarily prepared initial states. The Qiskit software development kit (SDK)\cite{Qiskit} is used to encode time evolution in quantum circuits, which are then run on the cloud-based IBM hardware. Time evolution of a single spin is simulated by mapping the spin to a qubit and applying a single unitary gate corresponding to the exact time-evolution operator. Results for two-spin and three-spin systems with nontrivial interactions are also presented. The scheme for performing time evolution on two-spin systems is exact, but it requires a more complicated circuit than the single-spin case. We are forced to use approximate methods when performing time evolution in systems with three or more spins. An additional challenge is that errors from currently-available devices make accurate results quite difficult to obtain for systems as small as three spins. However, we demonstrate how students can use Qiskit's built-in circuit simulator to explore systems with sizes up to $N=20$ spins. The circuit simulator is a fairly efficient and easy-to-use tool for exploring the expected output of quantum circuits for modest system sizes. Operationally, the Qiskit simulator consists of a classical computation which gives the expected output of a given quantum circuit subject only to statistical fluctuations. In this work we use the simulator to verify expected output of certain circuits. Using the simulator as a computational tool in its own right, we demonstrate evidence of a quantum phase transition that appears in a large spin system as a particular system parameter is varied. 

The remainder of this paper is organized as follows. Section~\ref{sec:s1} contains a review of relevant spin physics and the essential ingredients needed to simulate spin systems on IBM's quantum hardware. Dynamics in interacting systems of two and three spins are shown in Sec.~\ref{sec:s3}, and the Lie-Trotter decomposition is employed as a method for performing approximate time evolution in a three-spin system and on a large ($N=20$) system using Qiskit's circuit simulator. Finally, a brief discussion is included in Sec.~\ref{sec:s4}.

\section{Single spin dynamics}\label{sec:s1}
In this section we present the basic framework for simulating a spin degree of freedom on a quantum computer. To illustrate the approach in the simplest possible setting, we discuss the interaction of a single spin with a static magnetic field. However, the methods detailed below can also accommodate time-dependent magnetic fields with only minor modifications.~\cite{Alves} 

A single spin-$\frac{1}{2}$ degree of freedom is a two-level system whose quantum state can be expressed as a linear combination of basis kets $\left|\psi\right\rangle = \alpha\left|+\right\rangle + \beta \left|-\right\rangle$, where $\left|\pm\right\rangle$ are eigenstates of the operator $\hat{S}^{z}$, satisfying
\begin{eqnarray}
\hat{S}^{z}\left|\pm\right\rangle & = & \pm \frac{\hbar}{2}\left|\pm\right\rangle.\label{eq:eigz}
\end{eqnarray}
We follow standard notation for labeling the eigenstates of $\hat{S}^{z}$ (e.g., Ref.~\onlinecite{McIntyre}), but we caution the reader with experience in quantum information theory that these labels often refer to eigenstates of $\hat{S}^{x}$ in other sources.\cite{QiskitBook} The complex amplitudes are constrained by the normalization of the quantum state, $\left\langle \psi |\psi\right\rangle = |\alpha|^{2} + |\beta|^{2} = 1$. As the global phase of $\left|\psi\right\rangle$ is unobservable, a common parameterization is given by $\alpha = \cos\frac{\theta}{2}$, $\beta = e^{i\phi}\sin\frac{\theta}{2}$. To aid in calculations, the basis states can be represented by $|+\rangle \dot{=} (1,0)^{\mbox{\scriptsize T}}$ and $|-\rangle \dot{=} (0,1)^{\mbox{\scriptsize T}}$ so that the state $\left|\psi\right\rangle\rightarrow|\psi(\theta,\phi)\rangle$ can represented by the two-component vector
\begin{eqnarray}
\left|\psi(\theta,\phi)\right\rangle & \dot{=} & \left(\begin{array}{c} \cos\frac{\theta}{2}\\ e^{i\phi}\sin\frac{\theta}{2}\end{array}\right).\label{eq:state}
\end{eqnarray}
By employing the explicit matrix representations of the spin operators $\hat{S}^{\alpha} = \frac{\hbar}{2}\hat{\sigma}^{\alpha}$, where
\begin{eqnarray}
\hat{\sigma}^{x} &  \dot{=} & \left(\begin{array}{cc} 0 & 1 \\ 1 & 0 \end{array}\right), \;\;\;\; \hat{\sigma}^{y} \;\;\dot{=}\;\; \left(\begin{array}{cc} 0 & -i\\ i & 0 \end{array}\right),\nonumber\\
\hat{\sigma}^{z} & \dot{=} &\left(\begin{array}{cc} 1& 0 \\ 0 & -1 \end{array}\right),\label{eq:spinop}
\end{eqnarray}
one may compute the expectation values of the spin operators $\langle \hat{S}^{\alpha}\rangle  \equiv \langle \psi | \hat{S}^{\alpha} | \psi\rangle$ for the state in Eq.~(\ref{eq:state}). Simplifying with standard, trigonometric identities, we obtain
\begin{eqnarray}
\left\langle \hat{S}^{x} \right\rangle = \frac{\hbar}{2}\sin\theta\cos\phi,\nonumber\\
\left\langle \hat{S}^{y} \right\rangle = \frac{\hbar}{2}\sin\theta\sin\phi,\nonumber\\
\left\langle \hat{S}^{z} \right\rangle = \frac{\hbar}{2}\cos\theta.
\end{eqnarray}
In this form, it is apparent that the parameters $\theta$ and $\phi$ correspond to angular coordinates which map the possible states to points on an abstract, unit sphere---also known as the Bloch sphere.\cite{MikeIke} Time evolution for a single spin is generated by the interaction of the spin with a magnetic field ${\bf B}$ via the Hamiltonian
\begin{eqnarray}
\hat{H}_{1} = -\mu {\bf B}\cdot \hat{\bf S},\label{eq:ham}
\end{eqnarray}
where the constant $\mu$ represents the magnetic moment which couples the spin to an external field. The time-dependent state $\left|\psi(t)\right\rangle$ is obtained by solving the Schr\"{o}dinger equation,
\begin{eqnarray}
i\hbar\frac{d}{dt}\left|\psi(t)\right\rangle & = & \hat{H}\left|\psi(t)\right\rangle,
\end{eqnarray}
given some initial state $\left|\psi(0)\right\rangle$ and appropriate Hamiltonian operator, $\hat{H}$. The calculation of $\left|\psi(t)\right\rangle$ is typically performed by representing $\hat{H}$ as a $2\times 2$ matrix and extracting separate, coupled, first-order equations for the time-dependent amplitudes $\alpha(t)$ and $\beta(t)$, which can be solved using standard methods.\cite{GriffithsQM} With the aim of simulating such dynamics on a quantum computer, it is convenient to take a different approach. One may write the formal solution as
\begin{eqnarray}
\left|\psi(t)\right\rangle = \exp\left[-\frac{i}{\hbar}\hat{H}t\right]\left|\psi(0)\right\rangle \equiv \hat{U}(t)\left|\psi(0)\right\rangle,\label{eq:psit}
\end{eqnarray} 
and calculate the time-evolution operator $\hat{U}(t)$ explicitly as a $2\times 2$ matrix. As matrix exponentiation is not always covered in introductory quantum mechanics, we consider the representative example $\hat{H} = \hat{H}_{1}$ with ${\bf B} = B_{0}\hat{\bf x}$ in detail. In this case, the time-evolution operator can be written
\begin{eqnarray}
\hat{U}(t) & = & \exp\left[\frac{i\omega_{0}t}{2}\hat{\sigma}^{x}\right],\label{eq:ubx}
\end{eqnarray}
where $\omega_{0}\equiv \mu B_{0}$. Using the property that $\left[\hat{\sigma}^{x}\right]^{2} = \hat{I}$, one may expand the exponential as a Taylor series, then group even and odd terms separately to obtain,
\begin{eqnarray}
\hat{U}(t) & = & \hat{I}\left(1-\frac{(\omega_{0}t/2)^{2}}{2!} + \frac{(\omega_{0}t/2)^{4}}{4!} + \cdots\right) \nonumber\\
& & + i\hat{\sigma}^{x}\left(\omega_{0}t/2 - \frac{(\omega_{0}t/2)^{3}}{3!} + \frac{(\omega_{0}t/2)^{5}}{5!} + \cdots\right)\nonumber\\
& = & \cos\frac{\omega_{0}t}{2}\hat{I} + i\sin\frac{\omega_{0}t}{2}\hat{\sigma}^{x} \nonumber\\
& \dot{=} & \left(\begin{array}{cc} \cos\frac{\omega_{0}t}{2} & i\sin\frac{\omega_{0}t}{2}\\i\sin\frac{\omega_{0}t}{2} & \cos\frac{\omega_{0}t}{2}\end{array}\right).\label{eq:exph}
\end{eqnarray}
Here we have made use of the Taylor expansions for $\sin\frac{\omega_{0}t}{2}$ and $\cos\frac{\omega_{0}t}{2}$ to sum the infinite series explicitly. For the specific initial state $\left|\psi(0)\right\rangle = \left|+\right\rangle$, one can employ Eqs.~(\ref{eq:state})--(\ref{eq:spinop}) to obtain
\begin{eqnarray}
\left\langle \hat{S}^{x}(t)\right\rangle & = & 0,\nonumber\\
\left\langle \hat{S}^{y}(t)\right\rangle & = & \frac{\hbar}{2}\sin\omega_{0}t ,\nonumber\\
\left\langle \hat{S}^{z}(t)\right\rangle & = & \frac{\hbar}{2}\cos\omega_{0}t.\label{eq:spinex}
\end{eqnarray}
One might recognize these final expressions as a description of the steady Larmor precession exhibited by a spin in a static magnetic field. Larmor precession lies at the heart of nuclear magnetic resonance (NMR) and magnetic resonance imaging (MRI) and is discussed in several widely-used texts on quantum mechanics.\cite{McIntyre,GriffithsQM} Notably, the precession frequency $\omega_{0}$ is independent of the initial spin orientation, only depending on field strength $B_{0}$. 

Equations~(\ref{eq:spinex}) are the types of theoretical predictions we wish to test on a quantum computer.  In a quantum computer, the fundamental object is a qubit which can exist in a superposition of two basis states. These abstract basis states are typically labeled $|0\rangle$ and $|1\rangle$. Using the widely-adopted, ``computational basis,'' these basis states take the representations, $|0 \rangle\dot{=} (1,0)^{\mbox{\scriptsize T}}$, $|1 \rangle\dot{=} (0,1)^{\mbox{\scriptsize T}}$. Thus, a quantum spin is normally represented by a qubit with the mapping
\begin{eqnarray}
|+\rangle & \rightarrow & |0\rangle,\nonumber\\
|-\rangle & \rightarrow & |1\rangle.\label{eq:cbasis}
\end{eqnarray}
Admittedly, the opposite mapping might reasonably appear more natural based on the labels. But using Eqs.~(\ref{eq:cbasis}), the matrix representations of Pauli operators in the $\hat{S}^{z}$ eigenbasis will be identical to those in the computational basis, greatly simplifying the design of quantum circuits to simulate spin physics.

Users may access quantum hardware through the Qiskit SDK.~\cite{Qiskit} The Qiskit textbook~\cite{QiskitBook} and the IBM Quantum Learning resources~\cite{IBMcourse} provide detailed information on using Qiskit to construct quantum circuits, so we only sketch the main steps in translating the physical problem of spin dynamics into a quantum circuit. The reader is referred to other excellent resources~\cite{Brody,QiskitBook,IBMcourse} for additional background information. Sample Jupyter notebooks containing programs used in this work are included as supplementary material.

Universal quantum computers execute circuits of series of gates that modify the states of the qubits. To extract information about the system, a measurement must be performed. Thus, the three basic stages in a quantum circuit appropriate for simulating the spin dynamics discussed above are: (1) state initialization, (2) time evolution of the state, and (3) measurement of the spin components. IBM's quantum hardware automatically initializes each qubit to the state $\left|0\right\rangle$. An arbitrary quantum state of the form in Eq.~(\ref{eq:state}) can be generated by applying the following unitary operator,
\begin{eqnarray}
\hat{U}_{3}(\theta,\phi,\lambda) & \dot{=} & \left(\begin{array}{cc} \cos\frac{\theta}{2} & -e^{i\lambda}\sin\frac{\theta}{2}\\ e^{i\phi}\sin\frac{\theta}{2} & e^{i(\phi + \lambda)}\cos\frac{\theta}{2}\end{array}\right).\label{eq:ugate}
\end{eqnarray}
This $\hat{U}_{3}$ gate is available as a three-parameter gate in Qiskit, \texttt{u(theta,phi,lambda)}. The operator in Eq.~(\ref{eq:ugate}) is actually more general than we require to create the state in Eq.~(\ref{eq:state}). Moreover, the careful reader might observe that $\hat{U}_{3}(\theta,\lambda,\phi)$ is not Hermitian for arbitrary choices of $\theta$, $\phi$, $\lambda$. Operators corresponding to physical observables must be Hermitian, but unitarity ($\hat{O}\hat{O}^{\dagger} = \hat{O}^{\dagger}\hat{O} = \hat{I}$) is the only restriction on operators which change the state of a quantum system. This operation provides an efficient mechanism for generating a qubit in an arbitrary state $|\psi(\theta,\phi) \rangle = \hat{U}_{3}(\theta,\phi,0)|0\rangle$. Conveniently, time evolution is performed by noting that our expression for the time evolution operator $\hat{U}(t)$ in Eq.~(\ref{eq:ubx}) may also be written in terms of the $\hat{U}_{3}$ gate as $\hat{U}(t) = \hat{U}_{3}\left(\omega_{0}t,+\frac{\pi}{2},-\frac{\pi}{2}\right)$.




When using IBM quantum devices, measurements can only be performed in the computational basis. That is, for a qubit in the state $|\psi\rangle = \alpha |0\rangle + \beta |1\rangle$, a measurement will only return 0 or 1 corresponding to the measured state being $|0\rangle$ or $|1\rangle$, respectively. This constraint suggests that a single measurement operation can only measure $\hat{S}^{z}$. Measurement of spin projection along other directions is made possible by rotating the system appropriately before making a measurement in the computational basis.\cite{Alves,Brody} Within Qiskit, one may perform such an operation through the use of rotation gates, $\hat{R}_{\alpha}(\theta)$, for $\alpha = x,y,z$. With ordinary (classical) three-dimensional vectors in physical space, a rotation is performed by applying a standard rotation matrix. For example, to rotate a three-dimensional vector by an angle $\theta$ about the $x$-axis, the appropriate transformation is given by
\begin{eqnarray}
 {\bf R}_{x}(\theta) & = & \left(\begin{array}{ccc} 1 & 0 & 0 \\ 0 & \cos\theta & -\sin\theta\\ 0 & \sin\theta & \cos\theta\end{array}\right).\label{eq:rotmat}
 \end{eqnarray}
For quantum mechanical systems, rotations can be more subtle. While we associate spin components with directions in real space, the two-component object representing a spin exists in the abstract, Bloch sphere rather than in real space. Clearly, the three-dimensional rotation matrices cannot be applied to such a state. Rotation of quantum states is not commonly included in a traditional undergraduate course on quantum mechanics, so we quote the needed result and demonstrate its plausibility. Suppose we take a state $|\psi\rangle$ and perform a rotation about the $\alpha$-axis ($\alpha = x,y,z$) by angle $\theta$, transforming the state to $|\psi'\rangle = \hat{R}_{\alpha}(\theta)|\psi\rangle$. The appropriate operator, derivable from commutation relations of the spin-$\frac{1}{2}$ operators, is $\hat{R}_{\alpha}(\theta) = \exp\left[-i\hat{\sigma}^{\alpha}\theta/2\right]$.\cite{MikeIke,Sakurai} The diligent reader can use Eqs.~(\ref{eq:spinop}) and the method of matrix exponentiation discussed above to show
\begin{eqnarray}
\hat{R}_{\alpha}(\theta) & = & \hat{I}\cos\frac{\theta}{2} - i\hat{\sigma}^{\alpha}\sin\frac{\theta}{2}.
\end{eqnarray}
If we compute the expectation value of, say, $\hat{S}^{z}$ in a rotated state $|\psi'\rangle = \hat{R}_{x}(\theta)|\psi\rangle$, we get
\begin{eqnarray}
\langle \psi '|\hat{S}^{z}|\psi'\rangle & = & \langle \psi | \hat{R}_{x}^{\dagger}(\theta)\hat{S}^{z}\hat{R}_{x}(\theta)|\psi\rangle \nonumber\\
& = & \cos\theta \langle\psi|\hat{S}^{z}|\psi\rangle + \sin\theta\langle \psi|\hat{S}^{y}|\psi\rangle.\label{eq:rotspin}
\end{eqnarray}
The fastest way to arrive at Eq.~(\ref{eq:rotspin}) is to employ the commutation relation $\hat{\sigma}^{z}\hat{\sigma}^{x}-\hat{\sigma}^{x}\hat{\sigma}^{z}= 2i\hat{\sigma}^{y}$, the {\it anti}commutation relation $\hat{\sigma}^{x}\hat{\sigma}^{z} = -\hat{\sigma}^{z}\hat{\sigma}^{x}$, and the fact that $\left[\hat{\sigma}^{z}\right]^{2} = \hat{I}$. These properties are straightforward to verify with the explicit matrix representations of the Pauli operators in Eqs.~(\ref{eq:spinop}). Alternatively, it is also instructive to evaluate both sides of Eq.~(\ref{eq:rotspin}) by brute force for an arbitrary state $|\psi\rangle \dot{=} (\alpha,\beta)^{\mbox{\scriptsize T}}$. Regardless of overall strategy, double-angle formulas again prove essential to the simplification. 

To interpret Eq.~(\ref{eq:rotspin}), let $\langle \hat{O}\rangle \equiv \langle \psi | \hat{O} |\psi\rangle$ denote the expectation value of $\hat{O}$ in the original (unrotated) state. Then, consulting Eq.~(\ref{eq:rotmat}), the right-hand side of Eq.~(\ref{eq:rotspin}) represents the $z$-component of the vector formed by rotating of the vector $(\langle \hat{S}^{x}\rangle,\langle \hat{S}^{y}\rangle,\langle \hat{S}^{z}\rangle)^{\mbox{\scriptsize T}}$ about the $x$-axis by angle $\theta$. That is, rotating the state about the $x$-axis before measurement of $\hat{S}^{z}$ results in an effective spin measurement of the {\it un}rotated system about an axis $\hat{\bf n} =  \sin\theta\hat{\bf y}+\cos\theta\hat{\bf z}$. In summary, to measure spin about an arbitrary axis $\hat{\bf n}$, one can measure $\hat{S}^{z}$ after performing the rotation on the quantum state that corresponds to the classical rotation which brings $\hat{\bf n}$ into alignment with the $z$-axis. The reader is invited to check that ${\bf R}_{x}(\theta)\hat{\bf n} = \hat{\bf z}$.
 
\begin{figure}
\includegraphics[totalheight=4.5cm,]{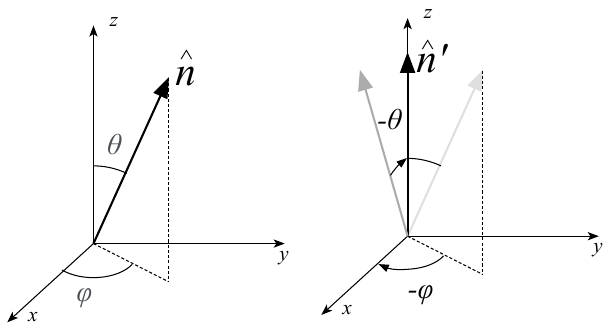}
\caption{To measure spin about the axis $\hat{\bf n}$, one must perform the required rotations to bring $\hat{\bf n}$ into alignment with the $z$-axis (the computational basis). Given polar-spherical angles $\theta$, $\phi$ corresponding to $\hat{\bf n}$, one finds $\hat{\bf n}' \equiv \hat{\bf z} = {\bf R}_{y}\left(-\theta\right){\bf R}_{z}\left(-\phi\right)\hat{\bf n}$.}
\label{fig:axes}
\end{figure}

Conveniently, rotation gates are accessible directly as predefined gates in Qiskit (e.g., \texttt{rx(theta)}). The motivated reader is encouraged to work out matrix representations for $\hat{R}_{\alpha}(\theta) = \exp\left[-i\hat{\sigma}^{\alpha}\theta/2\right]$ and verify that these non-Hermitian matrices are special cases of $\hat{U}_{3}(\theta,\lambda,\phi)$ for particular choices of $\theta$, $\lambda$, $\phi$. An algorithmic method for performing arbitrary rotations involving two rotations about fixed axes is depicted in Fig.~\ref{fig:axes}. It is also instructive to show explicitly that $\hat{U}_{3}(\theta,\phi,0) = e^{i\phi/2}\hat{R}_{z}(\phi)\hat{R}_{y}(\theta)$, so that one may view the unitary transformation $|\psi(\theta,\phi)\rangle = \hat{U}_{3}(\theta,\phi,0)|+\rangle$ in terms of the successive rotations which map $|+\rangle$ to any point on the Bloch sphere. The extra (global) phase $e^{i\phi/2}$ cannot be measured and has no physical effects. This equivalence is a particular example of the general statement that {\it any} unitary, single-qubit gate (or sequence of such gates) can be written as $\hat{U}_{3}(\theta,\phi,\lambda)$ for some choice of parameters $\theta$, $\phi$, $\lambda$.

Regarding measurements, we will need only to measure spin projections along the $\hat{\bf x}$ and $\hat{\bf y}$ directions. With this convention, these required rotations are encapsulated in the statement, $\hat{\bf z} = {\bf R}_{y}\left(-\frac{\pi}{2}\right)\hat{\bf x} = {\bf R}_{y}\left(-\frac{\pi}{2}\right){\bf R}_{z}\left(-\frac{\pi}{2}\right)\hat{\bf y}$. It is also possible to use the single rotation $\hat{\bf z} = {\bf R}_{x}\left(\frac{\pi}{2}\right)\hat{\bf y}$, but there are no significant gains in result quality for using different rotation ``formats.'' In practice, Qiskit performs an automatic, optimization procedure which simplifies a circuit as much as possible before execution on real hardware in order to reduce the effects of noise. We also note that if one were interested in the dynamics {\it after} such a measurement, it would be necessary to perform the corresponding inverse rotation after the measurement but before implementing additional time evolution. 

\begin{figure}[h!]
\includegraphics[scale=.7]{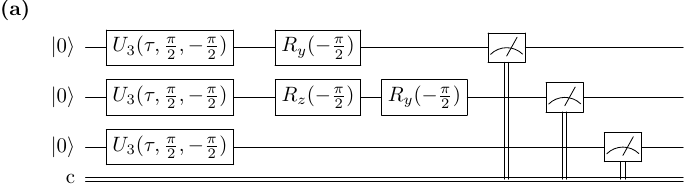} \\[3ex]
\includegraphics[scale=.7]{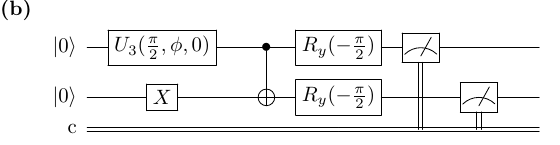}\\[3ex]
\includegraphics[scale=.7]{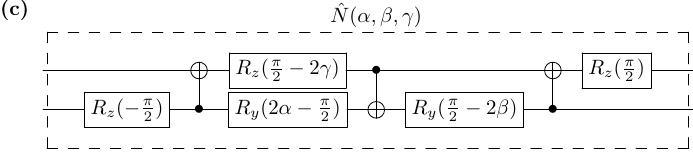}\\[3ex] 
\includegraphics[scale=.7]{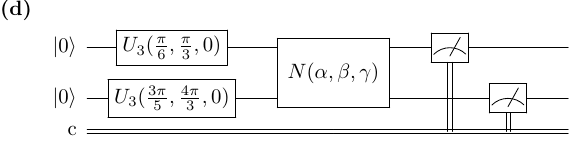}\\[3ex]
\includegraphics[scale=.7]{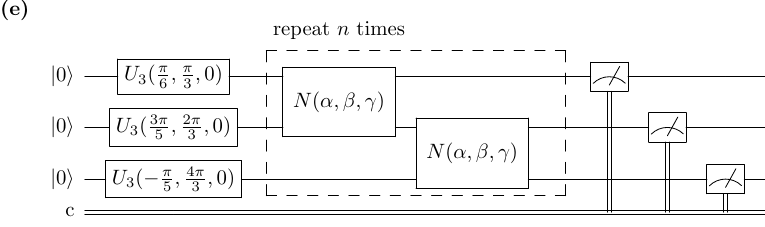}\\[3ex] 
\includegraphics[scale=.7]{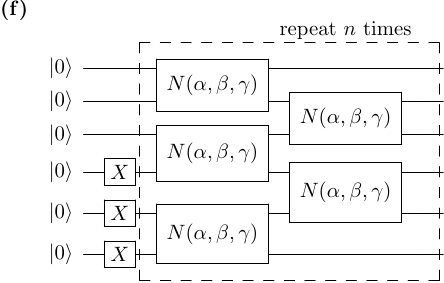}
\caption{Quantum circuits used to perform various tasks described in the main text; (a) circuit to perform time evolution and measure three different spin components for three copies of the same single spin; (b) circuit to create the two-spin state $|\psi(\phi)\rangle$ described in Sec.~\ref{sec:s3} and measure $\hat{S}^{x}$ on both spins; (c) circuit representation of $\hat{N}(\alpha,\beta,\gamma)$ defined and used to perform time evolution for two-spin system in Sec.~\ref{sec:s3}; (d) circuit to create two-spin state in Eq.~(\ref{eq:twospininit}) and perform time evolution under Hamiltonain $\hat{H}_{2}$ (Eq.~(\ref{eq:heisenberg})); (e) circuit for approximate time evolution of a three-spin state with Hamiltonian $\hat{H}_{N}$ (Eq.~(\ref{eq:xxz})); (f) circuit for initializing a domain wall state for $N=6$ spins and performing approximate time evolution.}\label{fig:circs}
\end{figure}

As IBM quantum devices are freely available, jobs sent from an open-access account are often queued---sometimes for significant times. With efficiency in mind, it is helpful to embed three independent, single-spin circuits into a single three-qubit circuit to measure all three spin components in a single job. The circuit shown in Fig.~\ref{fig:circs}a creates three copies of this single spin, performs time evolution generated by a single Hamiltonian on each spin, and then measures one of the three spin components on each qubit. Using Qiskit, the circuit in Fig.~\ref{fig:circs}a can be constructed and sent to IBM quantum devices for execution. 

When one executes the quantum circuit in Fig.~\ref{fig:circs}a, the result yields a sequence of three digits, each being either 0 or 1, which is written as a string to the classical register that uses classical bits to store the measurement results. For example, one might obtain the readout \texttt{`100'}, indicating that $+\frac{\hbar}{2}$ was obtained for $\hat{S}^{x}$ and $\hat{S}^{y}$, while $-\frac{\hbar}{2}$ was obtained for $\hat{S}^{z}$. The framework of quantum mechanics does not allow for exact predictions of the outcome of a single such experiment. We are only able to compute statistical quantities such as probabilities and expectation values, which give us information about the averages of many repetitions of an experiment performed on identically prepared systems. Typically, one executes a circuit for some large number of executions, or ``shots,''  so that experimental averages can be constructed and compared to theoretical predictions (e.g., Eqs.~\ref{eq:spinex}). With the release of Qiskit 1.x in 2024, circuit executions on IBM quantum hardware now require the use of Qiskit Runtime\cite{Runtime} environment. Within this framework, the user selects one of two possible ``primitives,'' or computational building blocks, for performing an experiment. Both primitives employ additional classical computing resources to optimize a particular job for a given device. The \texttt{Estimator} primitive returns estimates of expectation values for specified operators. This approach can be quite efficient for complicated circuits, but it obscures the actual manner in which these results are obtained. We will focus on the \texttt{Sampler} primitive, which is designed to return a (quasi)-probability distribution of the measurement outcomes. This approach has additional pedagogical value in requiring the user to reconstruct experimental averages from the low-level ``counts'' that accumulate over successive shots. These counts correspond to the number of times each state was observed during the repeated circuit executions. Examples of using the \texttt{Estimator} to calculate spin expectation values directly are contained in the supplementary material.

\begin{figure}
\includegraphics[totalheight=6.45cm,]{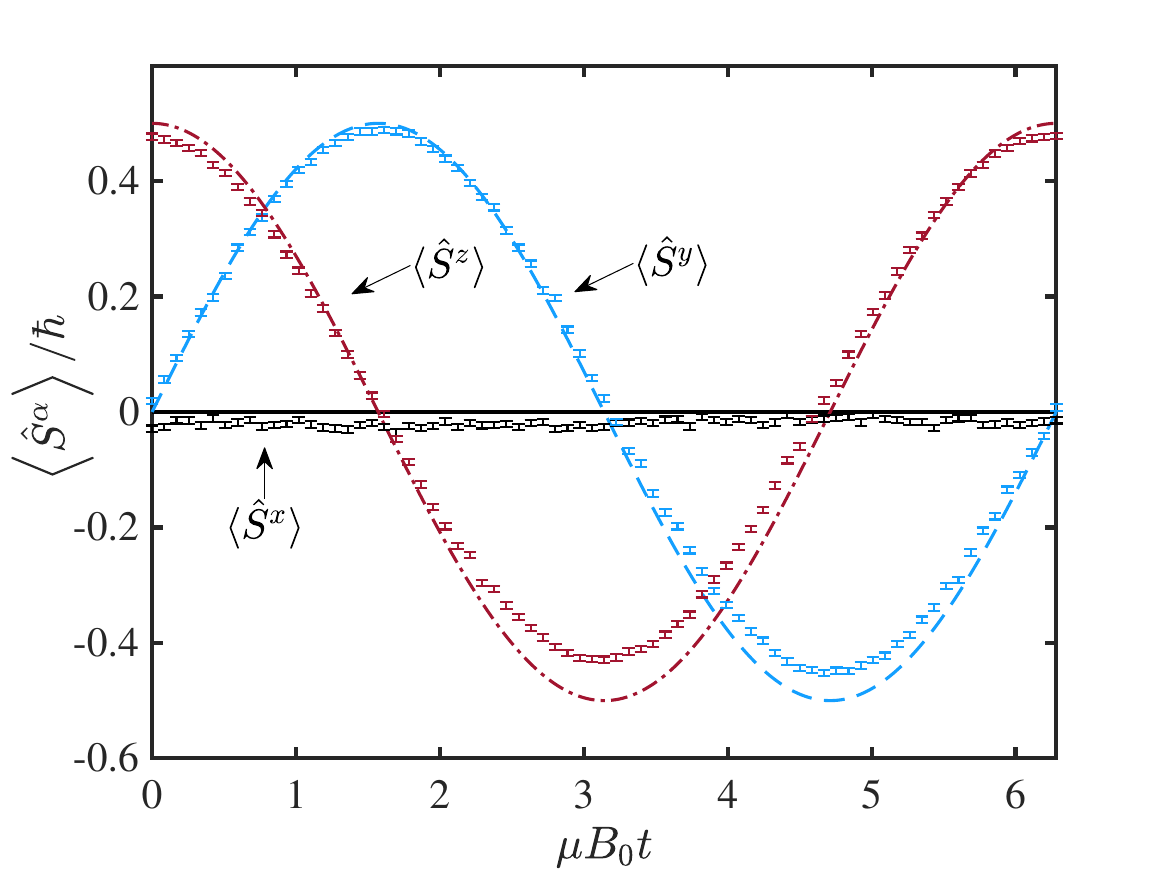}
\caption{Theory (lines) and actual results (markers with error bars) for time evolution of spin components with ${\bf B} = B_{0}\hat{\bf x}$ and $\left|\psi(0)\right\rangle = \left|+\right\rangle$ using circuit shown in Fig.~\ref{fig:circs}a on \texttt{ibmq\_casablanca} (v 1.2.35, a Falcon r4H processor).}
\label{fig:plot1}
\end{figure}

When executing a circuit using the \texttt{Sampler} primitive, the user can access the counts in the form of Python dictionary. This dictionary consists of pairs of variables, a bitstring (e.g., \texttt{`101'}) labeling each state, and a corresponding integer value representing the counts associated with that state. From this information, we construct spin expectation values as follows. Let $n_{q_{2}q_{1}q_{0}}$ be the counts for state $|q_{2}q_{1}q_{0}\rangle$, where $q_{0,1,2} \in \left\{0,1\right\}$. Taking $\langle  \hat{S}^{y}\rangle$ as an example, we have eight counts, $n_{000}, n_{001}, \cdots, n_{111}$. A weighted average of this spin component is given by
\begin{eqnarray}
\langle \hat{S}^{y}\rangle&  = & \frac{1}{N_{\mbox{\scriptsize shots}}}\left[\left(n_{000} + n_{001} + n_{100} + n_{101}\right)\left(+\frac{\hbar}{2}\right)\right.\nonumber\\
& &\left. + \left(n_{010} + n_{011} + n_{110} + n_{111}\right)\left(-\frac{\hbar}{2}\right)\right]
\end{eqnarray}
Note that the counts for which the middle qubit is measured to be $|0\rangle$ (corresponding to spin ``up'') weight the contribution of $+\frac{\hbar}{2}$, while those in which the middle qubit is in $|1\rangle$ (corresponding to spin ``down'') weight the contribution of $-\frac{\hbar}{2}$. Analogous averaging can be performed on the other qubits to obtain experimental values for the expected averages $\langle \hat{S}^{x}\rangle$, $\langle \hat{S}^{z}\rangle$. Everything discussed so far, from circuit construction to circuit execution, could be performed using the graphical user interface of the IBM Quantum Composer,\cite{Composer} accessible to anyone with a free IBM Quantum account. In that environment, results of circuit executions are returned as probability distributions, and it becomes cumbersome to perform any processing of count data for all but the simplest of circuits. Employing the Qiskit SDK within a Jupyter notebook (i.e., a Python environment) allows for straightforward extraction of the counts for computation of expectation values as described above. A more compelling reason for accessing IBM devices from within a Jupyter notebook is that it enables users to bundle much more complicated jobs into a single ``primitive unified bloc'' (PUB) to be sent to a quantum computer as a single unit. Since a quantum measurement affects the state of the system, the circuit in Fig.~\ref{fig:circs}a can only measure the spin components at a {\it single} time $t$. By making multiple copies of the circuit using a loop over a range of values for time $t$ or by employing built-in Qiskit techniques to rerun a circuit with different parameter values, one can obtain the spin measurements over a range of time samples. Results from running the circuit in Fig.~\ref{fig:circs}a on \texttt{ibmq\_casablanca} (v 1.2.35, a Falcon r4H processor) over a full cycle $0\leq t\leq 2\pi/\omega_{0}$ with 75 time samples and 8192 shots per sample are shown in Fig.~\ref{fig:plot1}. Error bars shown are statistical estimates given by $\sigma/\hbar \approx \frac{1}{2\sqrt{N_{\mbox{\scriptsize shots}}}}$.

The data in Fig.~\ref{fig:plot1} show better agreement with theory when $\langle \hat{S}^{\alpha}\rangle$ is closer to $+ \frac{\hbar}{2}$ than $-\frac{\hbar}{2}$. The simple reason for this uneven signal quality is rooted in the physical realization of the two qubit states, $|0\rangle$ and $|1\rangle$. IBM's devices employ transmon qubits~\cite{Gambetta} in which $|0\rangle$ represents the ground state of a small LC circuit (i.e., a quantum harmonic oscillator) with controllable anharmonicity due to a (nonlinear) Joesphson junction. This anharmonicity allows for isolation of the ground state and first excited state ($|1\rangle$) as the two states of a physical qubit. As $|1\rangle$ is an excited state, it decays to $|0\rangle$ with characteristic timescale $T_{1}\sim 100\,\mu$s, resulting in so-called ``spin-flip'' errors when the circuit's execution time becomes significant compared to $T_{1}$. Circuit execution time is linked to the time required to execute the circuit's gates rather than the physical time we are simulating. Thus, one expects higher rates of errors when the qubit state is placed in the state $|1\rangle$, which we have mapped to $|-\rangle$. 

The process of simulating time evolution on a quantum computer and constructing observables from the raw counts is an instructive process for students of quantum mechanics. Common student difficulties in interpreting expectation values, measurement, and the general mathematical structure of quantum mechanics are well documented.~\cite{Singh1,Singh2} The opportunity to construct operator expectation values from experimental measurements of common textbook-level systems gives students a different perspective of these abstract concepts. Though we have only considered the simplest case of Larmor precession, probing dynamics under time-dependent magnetic fields would require only modest modification to the basic scheme outlined in this section.~\cite{Alves} In the remainder of this paper, we shift focus to investigating the dynamics of multiple, interacting spins.

\section{Multiple spins}\label{sec:s3}
The main focus of this section is to explore the dynamics of interacting spins with quantum computers. Multi-particle quantum states give rise to new phenomena---such as entanglement---which are not present at the single-particle level. Before exploring the dynamics of multiple, interacting spins, we briefly examine some basic aspects of entangled spin states. We demonstrate that quantum computers can be used to distinguish between different types of entangled states by effectively measuring the total spin of the combined state. Additionally, this first exploration provides a simplified setting to introduce two-qubit gates which will be required for performing time evolution. Subsequent subsections treat dynamics of interacting spins for system sizes $N=2$ and $N=3$ separately. A final subsection outlines how to extend the approach for $N=3$ to larger systems with a specific focus on obtaining evidence for a phase transition in a particular spin model.
 
\subsection{Entanglement and spin}

The most general single-spin state (Eq.~(\ref{eq:state})) is represented by a two-component vector. For $N$ spins, a state is represented by a $2^{N}$-component vector with the most general state for $N=2$ taking the form
\begin{eqnarray}
|\Psi_{2}\rangle & \propto & c_{++}|+\rangle\otimes|+\rangle + c_{+-}|+\rangle\otimes|-\rangle\nonumber\\
& + & c_{-+}|-\rangle\otimes|+\rangle + c_{--}|-\rangle\otimes|-\rangle,
\end{eqnarray}
where $c_{\pm\pm}$ are complex constants. The symbol $\otimes$ denotes the tensor product operation. This construction reflects that a ``product'' of single-spin operators should act on a product of single-spin kets as $\left(\hat{A}\otimes \hat{B}\right)\left(|\psi_{1}\rangle \otimes |\psi_{2}\rangle\right) = \left(\hat{A}|\psi_{1}\rangle\right)\otimes \left(\hat{B}|\psi_{2}\rangle\right)$. In the same way, the notation $|-++\rangle$ for labeling a state of three spins in the previous section is actually shorthand for $|-\rangle\otimes|+\rangle \otimes |+\rangle$. Another common convention\cite{GriffithsQM} for writing the same state is $|-\rangle_{1}|+\rangle_{2}|+\rangle_{3}$, where site labels are included explicitly and the tensor product is ``hidden.'' We will not need an explicit representation of this operation in terms of matrices (e.g., the Kronecker product), so this notation merely provides a careful prescription for keeping track of the particular spins upon which an operator acts. The case of two spins provides simplest setting for probing quantum entanglement and quantum correlations. To this end, let us use the common shorthand (e.g., $|+-\rangle \equiv |+\rangle\otimes |-\rangle$) and consider the family of states defined by 
\begin{equation}
|\psi(\phi)\rangle = \frac{1}{\sqrt{2}}\left(|+-\rangle +e^{i\phi} |-+\rangle\right),\label{eq:psiphi}
\end{equation}
for some real phase angle $\phi$. This particular family of states is interesting because it interpolates smoothly between two particular, entangled states,
\begin{eqnarray}
|\Psi^{+}\rangle & = & |\psi(0)\rangle = \frac{1}{\sqrt{2}}\left(|+-\rangle + |-+\rangle\right),\nonumber\\
|\Psi^{-}\rangle & = & |\psi(\pi)\rangle = \frac{1}{\sqrt{2}}\left(|+-\rangle - |-+\rangle\right).
\end{eqnarray}
Saying that $|\Psi^{\pm}\rangle$ are entangled states is equivalent to the statement that there are no single-spin states $|\varphi_{1,2}\rangle$ for which one may write $|\Psi^{\pm}\rangle = |\varphi_{1}\rangle\otimes |\varphi_{2}\rangle$. That is, the state of one spin cannot be fully specified independently of the state of the other spin. Despite their strikingly similar structure, the states $|\Psi^{\pm}\rangle$ are physically distinct. Specifically, $|\Psi^{+}\rangle$ is a spin-1 state, and $|\Psi^{-}\rangle$ is a spin-0 state,\cite{GriffithsQM} and a measurement of total spin should distinguish between these two states. This is a particular example of the general statement that, while the overall phase of a state is not measurable, the relative phase of a superposition state (i.e., $e^{i0}$ versus $e^{i\pi}$) is measurable. 

Measuring the total spin of a two-spin system is a more complicated process than measuring single spin components, as described in the previous section. To represent a total-spin measurement in terms of single-spin properties that can be measured directly, let us write the total, squared-spin operator as
\begin{eqnarray}
\hat{S}_{\mbox{\scriptsize tot}}^{2} & = & \left(\hat{S}_{1}^{x}+\hat{S}_{2}^{x}\right)^{2} + \left(\hat{S}_{1}^{y}+\hat{S}_{2}^{y}\right)^{2} + \left(\hat{S}_{1}^{z}+\hat{S}_{2}^{z}\right)^{2}.\label{eq:spinsq}
\end{eqnarray}
Here we make use of the common convention $\hat{S}^{x}_{1} = \hat{S}^{x}\otimes\hat{I}$, $\hat{S}^{x}_{2} = \hat{I}\otimes\hat{S}^{x}$ for brevity. Squaring the first term yields quantities such as $[\hat{S}_{1}^{x}]^{2}$, $[\hat{S}_{2}^{x}]^{2}$, and $\hat{S}_{1}^{x}\hat{S}_{2}^{x}$. Using the tensor-product multiplication rule, 
\begin{eqnarray}
[\hat{S}_{1}^{x}]^{2} & = & [\hat{S}_{2}^{x}]^{2} = \frac{\hbar^{2}}{4}\hat{I}\otimes\hat{I},\nonumber\\
\hat{S}_{1}^{x}\hat{S}_{2}^{x} & = & \hat{S}^{x}\otimes \hat{S}^{x},
\end{eqnarray}
where $\hat{S}^{x}$ without label is the ordinary spin-$\frac{1}{2}$ operator $\frac{\hbar}{2}\hat{\sigma}^{x}$. Here we have employed the useful property of Eqs.~(\ref{eq:spinop}) that $[\hat{S}^{x}]^{2} = [\hat{S}^{y}]^{2} = [\hat{S}^{z}]^{2} = \frac{\hbar^{2}}{4}\hat{I}$. Collecting the remaining terms and taking an expectation value, we obtain
\begin{eqnarray}
\langle \hat{S}_{\mbox{\scriptsize tot}}^{2}  \rangle & = & \frac{3}{2}\hbar^{2} + 2\langle \hat{S}^{x}\otimes\hat{S}^{x} \rangle+ 2\langle\hat{S}^{y}\otimes\hat{S}^{y}\rangle + 2\langle\hat{S}^{z}\otimes\hat{S}^{z}\rangle,\nonumber\\\label{eq:stot}
\end{eqnarray}
which is valid for the expectation value taken with respect to any state. The quantities $\langle\hat{S}^{\alpha}\otimes\hat{S}^{\alpha}\rangle$ are sometimes called spin correlation functions and have been measured for various entangled, two-spin states in Ref.~\onlinecite{Brody}. Equation~(\ref{eq:stot}) demonstrates that we can effectively measure the expectation value of total (squared) spin for a two-spin state by measuring these spin correlation functions. For the entangled states $|\Psi^{\pm}\rangle$, each contribution of $\langle\hat{S}^{\alpha}\otimes\hat{S}^{\alpha}\rangle$ yields four terms which can be evaluated according to the rule $\left(\langle \varphi_{1}|\otimes\langle \varphi_{2}|\right)\hat{S}^{\alpha}\otimes\hat{S}^{\alpha}\left(|\varphi_{1}\rangle\otimes|\varphi_{2}\rangle\right) = \langle \varphi_{1}|\hat{S}^{\alpha}|\varphi_{1}\rangle\langle \varphi_{2}|\hat{S}^{\alpha}|\varphi_{2}\rangle$. It is also known from the general theory of spin angular momentum\cite{GriffithsQM} that eigenstates $|s\rangle$ of the spin operator in Eq.~(\ref{eq:spinsq}) satisfy
\begin{eqnarray}
\hat{S}^{2}_{\mbox{\scriptsize tot}}|s\rangle = \hbar^{2}s(s+1)|s\rangle,\label{eq:stotth}
\end{eqnarray}
where $s=0,\frac{1}{2},1,\cdots$ gives the maximal projection of spin on any axis in units of $\hbar$. What remains is to demonstrate that a quantum computer can effectively measure the appropriate values of $s$ for $|\Psi^{+}\rangle$ ($s=1$) and $|\Psi^{-}\rangle$ ($s=0$).

The state $|\psi(\phi)\rangle$ can be initialized for arbitrary $\phi$ by application of three, successive gates. First, one applies the a unitary gate to the first qubit to place it in a superposition state with a phase difference of $\phi$, 
\begin{eqnarray}
\hat{U}_{3}\left(\frac{\pi}{2},\phi,0\right) |00\rangle \rightarrow \frac{1}{\sqrt{2}}\left(|00\rangle + e^{i\phi}|10\rangle\right).
\end{eqnarray}
Next, applying a logical NOT operation $\hat{X} = \hat{U}_{3}(\pi,0,\pi)$ to the second qubit to yields the state $\frac{1}{\sqrt{2}}\left(|01\rangle + e^{i\phi}|11\rangle\right)$. Finally, one applies a controlled NOT (CNOT) operation with the first qubit as the control and the second qubit as the target. Specifically, a CNOT gates flips the target qubit {\it if} the control qubit is in the state $|1\rangle$. In the circuit diagram, the target qubit is denoted by the ``$+$'' and connected to the control qubit. The action of this gate on two qubit states can be summarized as
\begin{eqnarray}
\mbox{CNOT}\left[|0\rangle\otimes |0\rangle\right] & = & |0\rangle\otimes |0\rangle,\nonumber\\
\mbox{CNOT}\left[|0\rangle\otimes |1\rangle \right] & = & |1\rangle\otimes |1\rangle,\nonumber\\
\mbox{CNOT}\left[|1\rangle\otimes |0\rangle\right] & = & |1\rangle\otimes |0\rangle, \nonumber\\
 \mbox{CNOT}\left[|1\rangle\otimes |1\rangle \right] & = & |0\rangle\otimes |1\rangle,
\end{eqnarray}
with the qubits arranged in the form $|\mbox{target}\rangle\otimes|\mbox{control}\rangle$. 

Two-qubit gates such as CNOT typically lead to errors which are an order of magnitude larger than those associated with single-qubit gates.\cite{Smith} Consequently, it is desirable to use the minimum number of CNOT gates required for a given calculation, and we should expect larger errors in interacting systems than the single spin systems considered in Sec.~\ref{sec:s1}

\begin{figure}
\includegraphics[totalheight=6.45cm,]{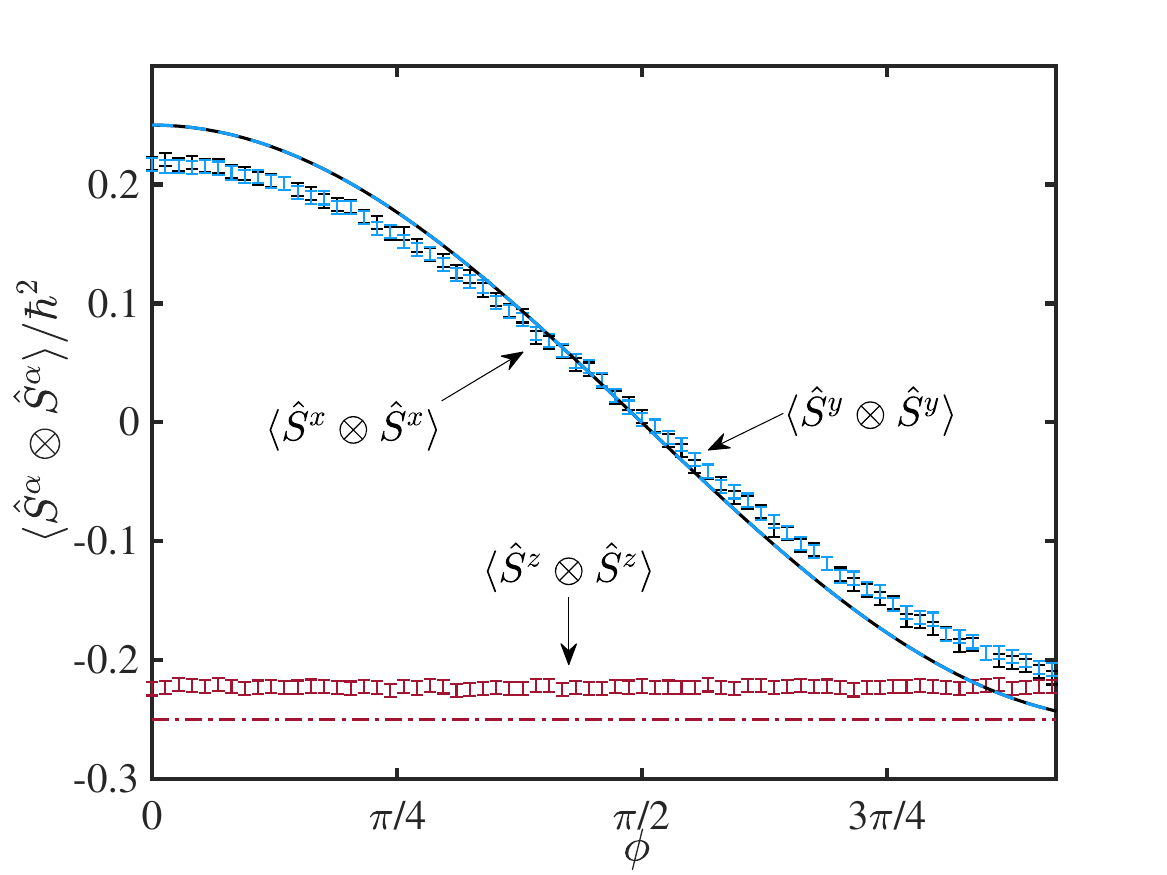}
\caption{Measurements of $\langle\hat{S}^{\alpha}\otimes\hat{S}^{\alpha}\rangle$ (markers with error bars) for $\alpha = x,y,z$ shown alongside theoretical predictions (lines) for the state $|\psi\rangle = (|+-\rangle + e^{i\phi}|-+\rangle)/\sqrt{2}$ with $0 \leq \phi \leq \pi$. The case $\phi=0$ corresponds to the $s=1$ state $|\Psi^{+}\rangle$, whereas $\phi = \pi$ corresponds to the $s=0$ state $|\Psi^{-}\rangle$. Data were collected from the device \texttt{ibm\_oslo} (v1.0.8, a Falcon r5.11H processor). Results for $\alpha=x,y$ are virtually indistinguishable, in agreement with theory.}
\label{fig:plota3}
\end{figure}

A circuit to prepare $|\psi(\phi)\rangle$ and measure $\langle\hat{S}^{x}\otimes\hat{S}^{x}\rangle$ is depicted in Fig.~\ref{fig:circs}b. As in the single-spin case, we cannot measure all three spin components with a single realization of the state. Thus we repeat the entire initialization, time evolution, and measurement process for the other spin components with 8192 shots per component measurement for each of the 75 samples of $\phi \in [0,\pi]$. Computing the spin correlation functions entails averaging the product of spin-components. For example, suppose execution of the circuit in Fig.~\ref{fig:circs}b yields $n_{00}$ counts of state $|00\rangle$, $\cdots$, $n_{11}$ counts for state $|11\rangle$. The corresponding spin correlation function is given by $\langle \hat{S}^{x}\otimes\hat{S}^{x}\rangle = \frac{\hbar^{2}}{4}(n_{00}+n_{11}-n_{01}-n_{10})/N_{\mbox{\scriptsize shots}}$. Results obtained from the device \texttt{ibm\_oslo} are shown in Fig.~\ref{fig:plota3}. The measured expectation value $\langle \hat{S}^{2}_{\mbox{\scriptsize tot}}\rangle$ is $(1.921\pm0.003)\hbar^{2}$ and $(0.189\pm0.003)\hbar^{2}$ for states $|\Psi^{+}\rangle$ and $|\Psi^{-}\rangle$, respectively. From Eq.~(\ref{eq:stotth}), these measurements should correspond $2\hbar^{2}$ and $0$, respectively. Due to the higher errors associated with CNOT gates, measured correlations on actual devices are typically smaller in magnitude than theoretical predictions when the expectation value is high and larger when the expectation value is low.\cite{Brody} However, these measurements are still roughly consistent with the predicted values, given the significant noise.


\subsection{Dynamics for $N=2$}
For the simplest case of $N=2$ interacting spins, a general two-body interaction Hamiltonian is
\begin{eqnarray}
\hat{H}_{2} = -J_{x}\hat{S}^{x}\otimes\hat{S}^{x} - J_{y}\hat{S}^{y}\otimes\hat{S}^{y} - J_{z}\hat{S}^{z}\otimes\hat{S}^{z},\label{eq:heisenberg}
\end{eqnarray}
where the $J_{x,y,z}$ are couplings. Additional, ``mixed'' terms such as $\hat{S}^{x}\otimes\hat{S}^{y}$ could also be included, but Eq.~(\ref{eq:heisenberg}) is sufficient to give rise to the rich dynamics discussed in this section. The structure of Eq.~(\ref{eq:heisenberg}) is motivated by the quantum mechanical exchange interactions~\cite{Giamarchi} which (for $J_{x,y,z}>0$) predicts a lower interaction energy for aligned magnetic dipoles (i.e., spins). For now, we take the couplings as arbitrarily tunable parameters to explore how to simulate such a system on a quantum computer. Later in this section, we return to the interesting question of how the coupling values affect the dynamics. 

\begin{figure}
\includegraphics[scale=0.81,left]{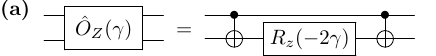}\\[5ex]
\includegraphics[scale=0.81,left]{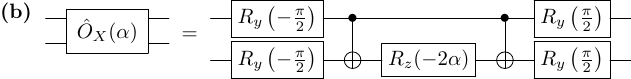}
\caption{(a) circuit representation of operator $\hat{O}_{Z}(\gamma)$; (b) circuit representation of $\hat{O}_{X}(\alpha)$.}
\label{fig:zzxx}
\end{figure}

To explore dynamics, we must compute the time evolution operator $\hat{U}(t) = \exp\left[-i\hat{H}t\right]$. For brevity, we employ units in which $\hbar\rightarrow 1$ for the remainder of this paper. For a Hamiltonian with the basic structure of Eq.~(\ref{eq:heisenberg}), we need a circuit representation of the operator $\hat{N}(\alpha,\beta,\gamma) \equiv \exp\left[i\alpha\hat{\sigma}^{x}\otimes\hat{\sigma}^{x} +i\beta\hat{\sigma}^{y}\otimes\hat{\sigma}^{y}  +i\gamma\hat{\sigma}^{z}\otimes\hat{\sigma}^{z}\right]$ for some constants $\alpha$, $\beta$, $\gamma$. Setting $\alpha = J_{x}/4$, $\beta = J_{y}/4$ and $\gamma = J_{z}/4$, the full time evolution operator $\hat{U}(t)$ is given conveniently by exactly $\hat{N}(\alpha, \beta, \gamma)$. To motivate the structure of a circuit representation for $\hat{N}(\alpha, \beta, \gamma)$, let us consider the simpler case of the operator
\begin{eqnarray}
\hat{O}_{Z} = \exp\left[i\gamma \hat{\sigma}^{z}\otimes\hat{\sigma}^{z}\right].
\end{eqnarray}
Working with four basis states, we wish to obtain a circuit that performs the following operations
\begin{eqnarray}
\hat{O}_{Z}|++\rangle & = & e^{i\gamma}|++\rangle,\nonumber\\
\hat{O}_{Z}|+-\rangle & = & e^{-i\gamma}|+-\rangle,\nonumber\\
\hat{O}_{Z}|-+\rangle & = & e^{-i\gamma}|-+\rangle,\nonumber\\
\hat{O}_{Z}|--\rangle & = & e^{i\gamma}|--\rangle.\label{eq:zzop}
\end{eqnarray}
Applying the circuit in Fig.~\ref{fig:zzxx}a to all four basis states using the matrix representation in the $\left\{|0\rangle, |1\rangle\right\}$ basis (or, equivalently, the $\left\{|\pm\rangle\right\}$ basis),
\begin{equation}
\hat{R}_{z}(\phi) = \exp\left[-i\phi\hat{\sigma}^{z}/2\right] \dot{=} \left(\begin{array}{cc} e^{-i\phi/2}  & 0 \\ 0 & e^{i\phi/2}\end{array}\right),
\end{equation}
one recovers the exact transformation defined by Eq.~(\ref{eq:zzop}). One could show this algebraically or explore experimentally by using the IBM Quantum Composer~\cite{Composer} to drag and drop circuit elements and explore the output for all four input basis states individually. The computational basis treats the $z$ direction preferentially, but simple modifications allow us to realize other operators such as $\hat{O}_{X}(\alpha) = \exp\left[i\alpha \hat{\sigma}^{x}\otimes \hat{\sigma}^{x}\right]$ and $\hat{O}_{Y}(\beta) = \exp\left[i\beta\hat{\sigma}^{y}\otimes \hat{\sigma}^{y}\right]$. One approach is to first rotate the state before performing the same operation and then to perform the opposite rotation to ``undo'' the initial rotation. For the case of $\hat{O}_{X}(\alpha)$, we can apply the methods of Sec.~\ref{sec:s1} to first rotate each qubit so that the $+x$ direction aligns with $|0\rangle$. Specifically, one can apply $\hat{ R}_{y}\left(-\frac{\pi}{2}\right)$ before the CNOT and $\hat{R}_{z}(-2\alpha)$ operations and then apply $\hat{ R}_{y}\left(+\frac{\pi}{2}\right)$, as shown in Fig.~\ref{fig:zzxx}b. 

Another set of carefully chosen rotations also yields a circuit representation for $\hat{O}_{Y}(\beta)$. Unfortunately, obtaining the full time-evolution operator $\hat{N}(\alpha,\beta,\gamma)$ is {\it not} as simple as stitching these three circuits together. In general, $e^{\hat{A}+\hat{B}} \neq e^{\hat{A}}e^{\hat{B}}$ when $\hat{A}$ and $\hat{B}$ do not commute, and different spin-component operators at the same site do not commute. When using an approximation scheme (such as the Lie-Trotter decomposition discussed in Sec.~\ref{sec:s3}B), one only advances over small time steps, and this level of error might be acceptable. But in the case of two interacting spins, it is possible~\cite{Smith} to construct an exact representation for $\hat{N}(\alpha,\beta,\gamma)$ which is shown in Fig.~\ref{fig:circs}c. Note that this circuit, still consisting of only CNOT gates and rotations, requires only three CNOT gates instead of the {\it six} one would need to apply the sequence of circuits for $\hat{O}_{X}(\alpha)$, $\hat{O}_{Y}(\beta)$, and $\hat{O}_{Z}(\gamma)$. Vatan and Williams have shown~\cite{Vatan} that any two-qubit gate (e.g., $\hat{N}(\alpha,\beta,\gamma)$) can be constructed using no more than three CNOT gates, so this representation turns out to be optimal. We will now proceed to use this circuit to perform time evolution in two-spin systems.

\begin{figure}
\includegraphics[totalheight=4.0cm,]{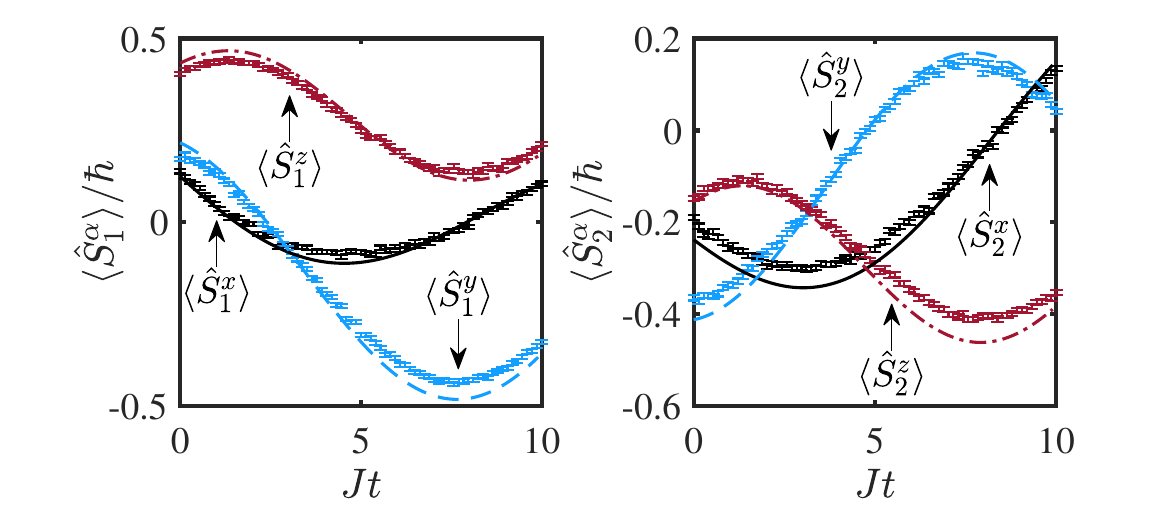}
\caption{Time-dependence of spin expectation values (markers with error bars) and theoretical predictions (lines) for an interacting, two-spin system using \texttt{ibmq\_bogota} to simulate the model in Eq.~(\ref{eq:heisenberg}) with the initial state given by Eq.~(\ref{eq:twospininit}).}
\label{fig:plot4}
\end{figure}

As a specific example, let us take $J_{x} = 0.5J$, $J_{y} = -0.45J$, $J_{z} = 0.25J$ for some arbitrary energy scale $J$ and simulate time evolution on an actual quantum computer. We take the following initial single-qubit states
\begin{eqnarray}
|\psi_{1}\rangle & = & \left(\cos\frac{\pi}{12}\right)|+\rangle + \left(e^{\frac{i\pi}{3}}\sin\frac{\pi}{12}\right)|-\rangle,\nonumber\\
|\psi_{2}\rangle & = & \left(\cos\frac{3\pi}{10}\right)|+\rangle + \left(e^{\frac{i4\pi}{3}}\sin\frac{3\pi}{10}\right)|-\rangle,\label{eq:twospininit}
\end{eqnarray}
which represent generic superposition states. Both the coupling parameters $J_{\alpha}$ and these initial spin states $|\psi_{1,2}\rangle$ are chosen to have no obvious spin structure and provide a representative example of generic dynamics. The initial two-spin state is $|\psi(0)\rangle = |\psi_{1}\rangle\otimes |\psi_{2}\rangle$. Dynamics generated by $\hat{H} = \hat{H}_{2}$ (Eq.~(\ref{eq:heisenberg})) with $J_{x} = 0.5J$, $J_{y} = -0.45J$, $J_{z} = 0.25J$, is simulated using the circuit shown in Fig.~\ref{fig:circs}d on the device \texttt{ibmq\_bogota} (v1.4.50, a Falcon r4L processor), which measures $\langle\hat{S}_{1,2}^{z}\rangle$ at a particular value of $t$. Two additional copies of this circuit with appropriate rotation gates for measuring the orthogonal spin-component expectation values are also used. With 8192 shots per time sample and 100 total time samples, each of the three circuits is executed a total of $100\times 8192$ times to generate the results shown in Fig.~\ref{fig:plot4}. The theoretical predictions are obtained by calculating $|\psi(t)\rangle = e^{-i\hat{H}_{2}t}|\psi(0)\rangle$ via an explicit exponentiation of the Hamiltonian matrix with the function \texttt{expm()} in \textsc{Matlab}.\cite{Matlab} A \textsc{Matlab} script which computes the theoretical predictions is included in the supplementary material.

In addition to computing spin expectation values, we can compute other observables directly from the counts used to produce Fig.~\ref{fig:plot4}. In particular, one may use the counts to compute basic spin correlation functions, $\langle \hat{S}^{\alpha}\otimes\hat{S}^{\beta}\rangle$ for $\alpha,\beta\in \left\{x,y,z\right\}$, as described in the previous subsection. Such correlations give us directly the expectation value of $\hat{H}$, since
\begin{equation}
\langle \hat{H}\rangle = -J_{x}\langle\hat{S}^{x}\otimes\hat{S}^{x}\rangle -J_{y}\langle\hat{S}^{y}\otimes\hat{S}^{y}\rangle - J_{z}\langle\hat{S}^{z}\otimes\hat{S}^{z}\rangle.
\end{equation}

\begin{figure}
\includegraphics[totalheight=4.0cm,]{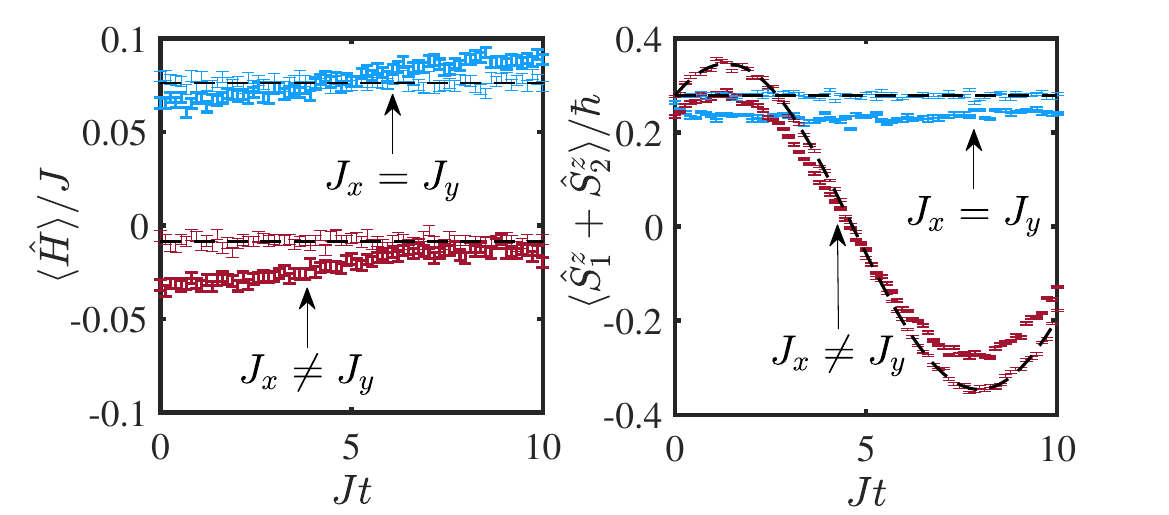}
\caption{Expectation value of energy, $\langle\hat{H}\rangle$, (left) and total $z$-projection of spin, $\langle\hat{S}_{1}^{z} + \hat{S}_{2}^{z}\rangle$, (right) as functions of time for the initial state in Eq.~(\ref{eq:twospininit}) for two sets of parameters ($J_{x}=J_{y}$ and $J_{x}\neq J_{y}$). For $J_{x}=J_{y}$, both energy and $\langle \hat{S}_{1}^{z}+\hat{S}_{2}^{z}\rangle$ should be conserved, whereas only energy is conserved for $J_{x}\neq J_{y}$. Data (markers with error bars) were collected using \texttt{ibm\_lagos} (v1.0.30, a Falcon r5.11H processor). Theoretical predictions (lines) were obtained via exact diagonalization. Circuit simulator results (faint markers) are in good agreement with theory.}
\label{fig:plota2}
\end{figure}

In Fig.~\ref{fig:plota2}, we show the time evolution of $\langle \hat{H}\rangle$ and $\langle\hat{S}_{1}^{z} + \hat{S}_{2}^{z}\rangle$ for two different sets of parameters in Eq.~(\ref{eq:heisenberg}).  As the Hamiltonian $\hat{H}_{2}$ (Eq.~(\ref{eq:heisenberg})) describes a closed system with no explicit time dependence (e.g., $\hat{H}_{2}$ possesses time-translation invariance), $\langle\hat{H}\rangle$ should be constant in time. This statement is an example of Noether's theorem,\cite{McIntyre} which states that conservation laws follow from symmetries. For $J_{x}=J_{y}$, rotational invariance of $\hat{H}_{2}$ (Eq.~(\ref{eq:heisenberg}) leads to the conservation of the total spin projection along the $z$ axis. Accordingly, one expects $\langle\hat{S}_{1}^{z} + \hat{S}_{2}^{z}\rangle$ to remain constant when $J_{x}=J_{y}$ but possibly vary when $J_{x}\neq J_{y}$. In both cases, $\langle \hat{H}\rangle$ should be constant in time. The initial state for both cases is given by Eq.~(\ref{eq:twospininit}). 

As noted previously the two-qubit, CNOT gates used to define the initial state and perform time evolution have significantly higher error rates than single qubit gates. These errors are compounded when computing averages of products of spin measurements.\cite{Brody} Accordingly, we find noticeable discrepancies between the actual results and the theoretical predictions in Fig.~\ref{fig:plota2}. For the symmetric case, we take $J_{x}=J_{y} = 0.5J$ and $J_{z} = 0.25J$. In this case, neither $\langle \hat{H}\rangle$ nor $\langle \hat{S}^{z}_{1} + \hat{S}^{z}_{2}\rangle$ exhibits significant dynamics, with each closely tracking its initial value. Breaking rotational symmetry by setting  $J_{y} = -0.45J \neq J_{x}$ results in conserved energy, but the total $z$ projection of spin varies significantly over the time scales considered. Qiskit also provides a circuit simulator that produces counts according to exact, theoretical results for a given circuit. The simulator is especially useful for the comparison of experimental results in order to assess the impact of the errors of the physical device. Simulator results (faint markers) are shown to be in good agreement with the theoretical predictions in Fig.~\ref{fig:plota2}, indicating that the circuit is performing the desired simulation but that the errors of the device are significant.

Sample Jupyter notebooks for obtaining the results and theoretical predictions in this section are included the supplementary material. In addition to exploring how parameter choices affect conservation laws, students could also generalize this basic framework to explore other types of systems. For example, in addition to the pure initial states considered in this work, it is possible to compute the dynamics of observables when the initial state is a mixed state, corresponding to thermodynamic equilibrium.~\cite{Doronin}

\subsection{Dynamics for $N=3$}

In this subsection, we outline how to investigate larger systems. For $N\geq 3$, it is generally not possible to perform exact time evolution. An exact representation of the time evolution in terms of quantum gates is generally possible only for models which can be diagonalized analytically and are therefore ``exactly solvable.'' Though such models do exist,~\cite{Cervera} the aim of this section is to equip the reader with tactics for the more general situation. The basic strategy, detailed below, uses the Lie-Trotter decomposition~\cite{Trotter} to break the full time evolution into small-time increments. Other, more efficient approximation schemes which scale better with increasing circuit depth are possible,~\cite{Kemper1} but the basic Lie-Trotter decomposition is chosen as a conceptually simple approach requiring minimal prerequisites to apply. It is also notable that the focus of the 2021 IBM Quantum Open Science Contest\cite{IBMopen} was to maximize state fidelity using a Lie-Trotter decomposition scheme to perform time evolution in a modest, three-qubit system.

To explore the method, let us consider a time-independent Hamiltonian which is composed of two parts that do not commute,
\begin{eqnarray}
\hat{H} & = & \hat{A} + \hat{B},
\end{eqnarray}
such that $\left[\hat{A},\hat{B}\right] \neq 0$. The time-evolution operator for this system is given by
\begin{eqnarray}
\hat{U}(t) = \exp\left[-i\hat{H}t\right] = \exp\left[-i\hat{A}t- i\hat{B}t\right].
\end{eqnarray}
For the case of $N=1,2$, we computed $\hat{U}(t)$ exactly. In general, such exact computations are not possible. Let us define the operators
\begin{eqnarray}
\hat{U}_{A}(t) & = & \exp\left[-i\hat{A}t\right],\;\;\;\;\;\; \hat{U}_{B}(t) = \exp\left[-i\hat{B}t\right].
\end{eqnarray}
As $\hat{A}$ and $\hat{B}$ do not commute, $\hat{U}(t) \neq \hat{U}_{A}(t)\hat{U}_{B}(t)$. However, according to the Lie-Trotter product formula,\cite{Trotter} one may write formally
\begin{eqnarray}
& &\exp\left[-i(\hat{A}+\hat{B})t\right] \nonumber\\& & = \lim_{n\rightarrow \infty}\left(\exp\left[-i\hat{A}t/n\right]\exp\left[-i\hat{B}t/n\right]\right)^{n},
\end{eqnarray}
or
\begin{eqnarray}
\hat{U}(t) & = & \lim_{n\rightarrow\infty}\left(\hat{U}_{A}\left(\frac{t}{n}\right)\hat{U}_{B}\left(\frac{t}{n}\right)\right)^{n}.\label{eq:trotterlimit}
\end{eqnarray}
That is, if we subdivide a time step into $n$ sub-steps, the product of exponentials will converge to the true expression for $\hat{U}(t)$ as $n\rightarrow\infty$. We note that including more Trotter steps requires more circuit gates, leading to an increased circuit depth. Since errors scale with circuit depth, there is a very real tradeoff between using more steps to aid in convergence of the algorithm and using as few steps as possible to minimize accumulation of errors. It has been shown\cite{Smith} that this optimization results in extremely limited timescales for accurate simulation of systems with size $N\sim\mathcal{O}(10)$.

Most IBM devices allow for the application of CNOT gates only between small subsets of nearest-neighboring qubits. Applying CNOT gates to qubits which are not directly connected is possible through the application of a sequence of so-called SWAP gates.\cite{QiskitBook} In the interest of minimizing circuit depth, we only consider nearest-neighbor interactions in small systems which can be mapped to physically-connected qubits. In this section, we consider the many-body ($N>2$) generalization of $\hat{H}_{2}$ (Eq.~(\ref{eq:heisenberg})) with only nearest-neighbor interactions,
\begin{eqnarray}
\hat{H}_{N} & = & \sum_{j=1}^{N-1}\left[-J_{x}\hat{S}_{j}^{x}\hat{S}_{j+1}^{x} - J_{y}\hat{S}_{j}^{y}\hat{S}_{j+1}^{y} - J_{z}\hat{S}_{j}^{z}\hat{S}_{j+1}^{z}\right].\label{eq:xxz}
\end{eqnarray} 
Here, we employ the shorthand $\hat{S}^{\alpha}_{j}\hat{S}_{j+1}^{\alpha} \equiv \hat{I}\otimes\cdots\otimes \hat{I}\otimes\hat{S}^{\alpha}\otimes\hat{S}^{\alpha}\otimes \hat{I}\otimes\cdots\otimes\hat{I}$. Such nearest-neighbor models are surprisingly useful tools for describing certain interacting, condensed-matter systems. To apply the Trotter decomposition to the time-evolution operator, let us define $\hat{h}_{jk} = -\left(J_{x}\hat{S}^{x}_{j}\hat{S}^{x}_{k} +J_{y}\hat{S}^{y}_{j}\hat{S}^{y}_{k} + J_{z}\hat{S}^{z}_{j}\hat{S}_{k}^{z}\right)\delta_{j+1,k},$ so that the time evolution operator can be written
\begin{equation}
\hat{U}(t) = \exp\left[-it\sum_{j,k}\hat{h}_{jk}\right] \approx \prod_{l=1}^{n}\prod_{j,k}e^{-i(t/n)\hat{h}_{jk}}.\label{eq:trotdecomp}
\end{equation}
Each factor of $e^{-i(t/n)\hat{h}_{jk}}$ corresponds to the circuit gate $\hat{N}(\alpha,\beta,\gamma)$ shown in Fig.~\ref{fig:circs}e, which contains three CNOT gates.

We consider three spins, each initialized to the state given in Eq.~(\ref{eq:state}) with $\theta_{1}=\frac{\pi}{6}$, $\phi_{1} = \frac{\pi}{3}$, $\theta_{2} = \frac{3\pi}{5}$, $\phi_{2} = \frac{4\pi}{3}$, $\theta_{3} = -\frac{\pi}{5}$, $\phi_{3} = \frac{2\pi}{3}$. These particular choices carry no particular significance and simply represent a typical state with no obvious structure. Time evolution is performed using the $N=3$ case of Eq.~(\ref{eq:xxz}). Students can use the programs in the supplementary material to explore how the results change with different initial states. 

\begin{figure*}[!ht]
\includegraphics[totalheight=5.1cm,]{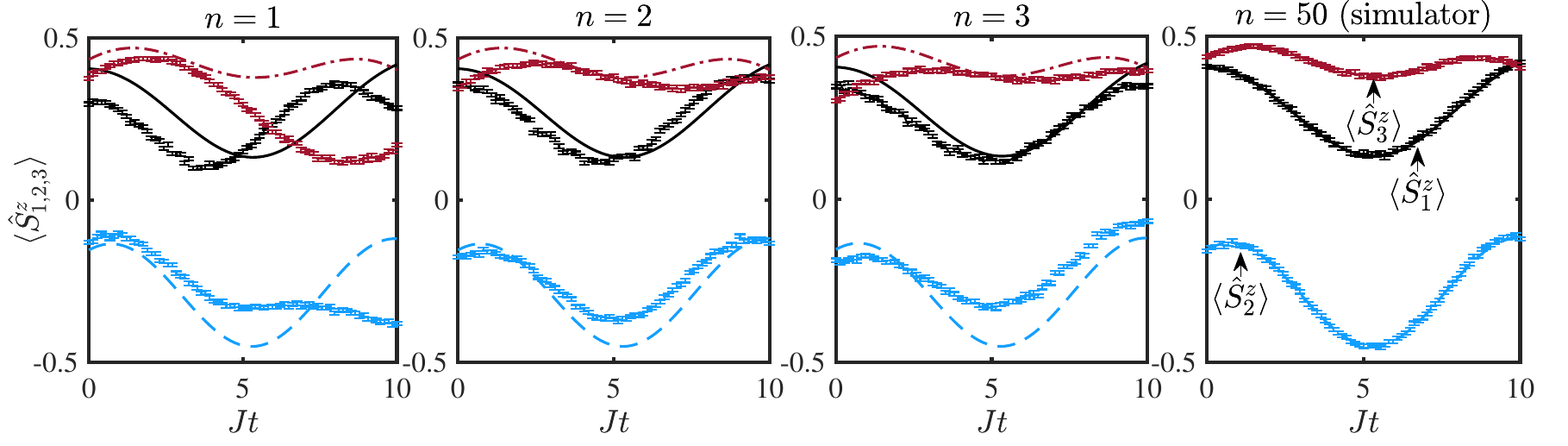}
\caption{Approximate time evolution of spin expectation values in three-spin system (Eq.~\ref{eq:xxz}) with variable number of Trotter steps using \texttt{ibm\_perth}. The initial state is given in the main text. Results from quantum hardware (markers with error bars) are shown with theory (lines). Results in the right-most panel $(n=50)$ are obtained from the circuit simulator rather than actual quantum hardware.}
\label{fig:plot5}
\end{figure*}

As with the $N=2$ case, we take $J_{x} = 0.5J$, $J_{y} = -0.45J$, $J_{z} = 0.25J$ and work with dimensionless time $Jt$. A schematic circuit for computing $\langle \hat{S}_{j}^{z}\rangle$ is shown in Fig.~\ref{fig:circs}e for $j=1,2,3$. The results obtained from \texttt{ibm\_perth} v1.1.14, a Falcon r5.11H processor, are depicted alongside theoretical predictions in Fig.~\ref{fig:plot5} for $n=1,2,3$ Trotter steps. To produce Fig.~\ref{fig:plot5}, the circuit in Fig.~\ref{fig:circs}e is executed for 8192 shots for each of the 75 time samples in the range $0\leq t \leq 10/J$. The challenge of simulating even a three-spin system is evident from the observation that $n=2$ yields notably better agreement between experiment and theory than $n=1$, but significant degradation occurs upon increasing the number of Trotter steps to $n=3$. The noiseless simulator results are unaffected by CNOT errors, as shown by the excellent agreement between simulation and theory when the number of Trotter steps is increased significantly to $n=50$. We note that the theoretical prediction is obtained from exact diagonalization without Trotter decomposition, so the right-most panel in Fig.~\ref{fig:plot5} demonstrates that the Trotter decomposition itself indeed converges to the exact result for a sufficiently large number of Trotter steps. A \textsc{Matlab} script to reproduce the theoretical predictions and a Jupyter notebook which executes the circuit in Fig.~\ref{fig:circs}e are included in the supplementary material. Due to the significant errors obtained for only three spins, investigation of larger systems is only possible by applying some form of error correction~\cite{Nation} or using a more sophisticated algorithm than the Lie-Trotter decomposition. 

\subsection{Many-body dynamics: spin chains}

The simple techniques of the preceding sections can now be applied to study complex, condensed-matter systems. We employ the quantum circuit simulator rather than actual hardware due to the significant errors associated with the significant number of Trotter steps required to study larger systems. In this section, we show results for the $XXZ$ spin chain, defined by the Hamiltonian $\hat{H}_{N}$ with $J_{x}=J_{y}\equiv J_{xy}$ and $N\gg1$. Giamarchi\cite{Giamarchi} describes the $XXZ$ model and its low-energy properties in the thermodynamic limit ($N\rightarrow\infty$) in detail. We will not repeat this full description here. Instead we give a brief summary and then discuss our results using the terminology of Giamarchi. We consider the antiferromagnetic regime in which $J_{z}<0$, while $J_{xy}>0$. In the thermodynamic limit, the $XXZ$ spin chain exhibits a quantum phase transition~\cite{Shankar,Sachdev} at the point $|J_{z}| = J_{xy}$. For $0<|J_{z}|<J_{xy}$, transport in the system is, ``ballistic,''\cite{Giamarchi} with excitations moving freely through the system without scattering. For $|J_{z}|>J_{xy}$, transport becomes ``diffusive,'' as excitations interact with each other via scattering processes.\cite{Sirker} 

We note that nonequilibrium {\it dynamics} in isolated, many-body systems is an active area of research.\cite{Polkovnikov} A particular topic of recent interest is ``quench dynamics'' in which many-body time evolution is computed for some arbitrary initial state which differs from the many-body ground state.\cite{Essler} The circuits described in the preceding sections effectively simulate quantum quenches for systems of size $N=1,2,3$. Here we consider system size $N=20$. To generate nonequilibrium dynamics, the initial state can be constructed as a ``domain wall'' spin configuration in which all spins in the left half the system are in the state $|+\rangle$, while those in the right half of the system are in $|-\rangle$. Succinctly, $|\Psi_{0}\rangle = |++\cdots+-\cdots --\rangle$. As the system evolves, the domain wall will spread. In the ballistic phase, it is expected that the domain edges will spread with constant speed. For larger $|J_{z}| > J_{xy}$, the transport should become diffusive.



\begin{figure*}[!ht]
\includegraphics[totalheight=7.8cm,]{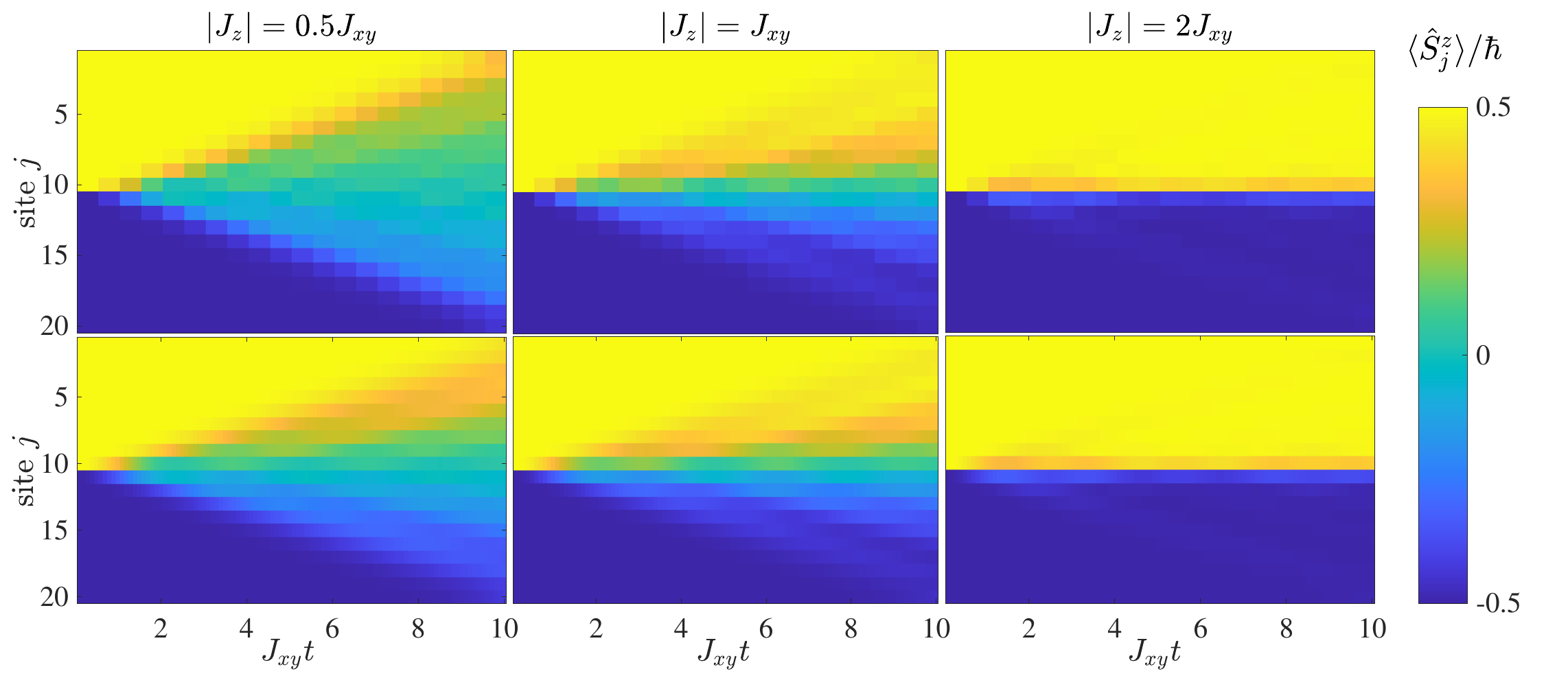}
\caption{(Color online) Magnetization dynamics $\langle \hat{S}_{j}^{z}\rangle$ are shown for time evolution of a domain wall initial state $|\Psi_{0}\rangle = |+\cdots+-\cdots -\rangle$ using the Hamiltonian $\hat{H}_{N}$ (Eq.~(\ref{eq:xxz})) with $J_{x} = J_{y} \equiv J_{xy}$ and various values of $J_{z}<0$. For $|J_{z}|<J_{xy}$, ballistic transport is observed in which the domain wall spreads at constant speed. A quantum critical point exists at $|J_{z}| = J_{xy}$, and transport becomes diffusive for $|J_{z}|>J_{xy}$. The cases considered are $J_{z} =- 0.5 J_{xy}$ (left), $J_{z} = -J_{xy}$ (center), and $J_{z} =- 2J_{xy}$ (right). Data collected from the quantum circuit simulator are shown on the top row, while corresponding theoretical predictions are shown in the bottom row. Quantum circuit simulator results were obtained with $n=100$ Trotter steps.}
\label{fig:plota1}
\end{figure*}

 A circuit to prepare the domain wall state for $N=6$ spins and perform approximate time evolution is shown in Fig.~\ref{fig:circs}f. Generalizing to larger systems such as $N=20$ is accomplished by implementing this pattern on a larger number of qubits. It is striking how the same structure used for $N=3$ spins (c.f., Fig.~\ref{fig:circs}e) generalizes immediately to larger systems with no significant, structural modifications. Results are shown in the top row of Fig.~\ref{fig:plota1} for $N=20$ spins, $8192$ shots, and $n = 100$ Trotter steps for each spin at each of the 20 time samples. For simplicity, we have used $n$ Trotter steps for each time sample. In practice~\cite{Smith}, one generally maintains constant Trotter step size $\Delta t$ and varies the number of Trotter steps according to simulation time $t$. For theoretical comparison, the bottom row of Fig.~\ref{fig:plota1} depicts predictions obtained using the \textsc{Expokit} package~\cite{Sidje} to compute the time-evolution operator from a sparse-matrix representation of Eq.~(\ref{eq:xxz}) in \textsc{Matlab}. A Jupyter notebook and a \textsc{Matlab} script which respectively reproduce the simulation and theoretical predictions are included in the supplementary material.

In general, excellent agreement is observed between simulation and theory. The transition from ballistic expansion of the domain wall to diffusive behavior in which the wall does not move appreciably at timescales considered is observed clearly. Somewhat crudely, one can think of the transition from ballistic expansion to diffusive spreading as being rooted in the strength of the $J_{z}$ term, which favors antiferromagnetic alignment in the $z$ direction for $J_{z}<0$. For $|J_{z}|>J_{xy}$, destroying the antiferromagnetic alignment of the two central spins in the domain wall presents an energy penalty which is not overcome by any order that develops in the $xy$-plane as the wall spreads. 

We note the potentially under-appreciated value of using the simulator to perform such calculations, especially for students. The sparse-matrix-based approach using \textsc{Expokit} is quite efficient, producing each panel of Fig.~\ref{fig:plota1} with 100 time samples in approximately ten minutes. Each panel on the top row required roughly an hour of computation time on the same personal computer for 20 time samples. But even with a simple-to-use software package such as \textsc{Expokit}, the user must be able to write a program from scratch to represent the Hamiltonian in Eq.~(\ref{eq:xxz}) and observables (e.g, $\hat{S}_{j}^{z}$) as sparse matrices. It is notable that the same thermodynamic-limit physics (ballistic vs. diffusive transport) can be probed using a circuit scarcely more complex than that used for the system size $N=3$. The application to $N\gg3$ requires a many more Trotter steps to converge than the $N=3$ case. But this is not a fundamental change in the complexity of the circuit's working principle, as additional Trotter steps can be implemented with a loop. Thus, the barrier to entry for students and non-experts to use the quantum circuit simulator to investigate many-body physics is arguably lower than that of many traditional computational approaches. Admittedly, the landscape of user-friendly, high-level libraries for probing these types of many-body systems is evolving rapidly. Some recently-developed libraries for computing approximate time evolution in many-body systems are quite easy to implement and undeniably much more efficient than quantum circuit simulation. Notably, the iTensor\cite{itensor} library allows users to compute approximate dynamics using matrix-product state (MPS) tensor networks for systems with $N\sim \mathcal{O}(100)$. However, we argue that there is significant pedagogical value to developing a fairly robust many-body simulation from scratch.

\section{Discussion}\label{sec:s4}

We have presented schemes for simulating small ($N\leq 3$) spin systems on IBM quantum computers and measuring the time-dependent expectation values of various spin components. Systems of size $N=1,2$ have been treated exactly, and the results contained only modest errors due to the simple circuits employed. For larger systems, approximate schemes must be employed to perform time evolution. A Lie-Trotter decomposition was used to approximate the time-evolution operator in a three-spin system. Even in a system with as few as three spins, the tradeoff between adding circuit elements to yield convergence of the algorithm and reducing circuit elements to minimize noise becomes apparent. Finally, Qiskit's quantum circuit simulator has been shown to be a useful and easy-to-use tool for simulating large ($N=20$) interacting quantum systems.

The dynamics of small, spin systems are treated frequently in introductory quantum mechanics. The use of free, cloud-based IBM hardware provides a highly accessible opportunity for students to explore these systems experimentally. When comparing experimental results to theoretical predictions, students gain valuable experience in extracting experimental values of physical observables by manipulating the raw counts which are returned from the quantum circuit executions. Students must wrestle with what it means to ``measure'' a quantum mechanical observable, averaging a large number of independent results, and compute statistical uncertainties. Extensions of the problems presented here could form the basis of student projects in which the basic tools are applied to more complex systems (e.g., assessing effects of next-nearest neighbor interactions in larger systems using the quantum circuit simulator, or investigating time-dependent systems). Using the approach detailed in this work, students are well-equipped to use quantum computers and quantum circuit simulators to probe interesting physics.

In addition to providing a useful and accessible experimental component to the traditional undergraduate treatment of quantum mechanics, these types of simulations give students experience using cutting-edge technology. In this sense, the seemingly-abstract content in a quantum mechanics course has direct relevance to an exciting area of industry. Companies such as IBM,\cite{IBMQ}, Rigetti,\cite{Rigetti} and IonQ,\cite{IonQ} have devoted considerable resources to constructing functional NISQ devices. Such devices represent a small-but-important step toward the construction of fault-tolerant quantum computers. The current generation of hardware is far from capable of breaking RSA encryption, simulating complex molecules for drug design, or performing any of the other tasks which likely motivate large corporations to invest heavily in the technology. But this currently-available hardware is capable of simulating---albeit {\it imperfectly}---the type of conceptually simple quantum physics one finds in undergraduate-level textbooks on quantum mechanics. Thus, students with a working knowledge of basic quantum mechanics are well equipped to explore such technology. It is not yet clear if sufficiently clever error correction and programming will enable NISQ devices to show a legitimate quantum advantage over classical computers or if they simply represent a pivotal step in the evolution of practical quantum computation. Excellent resources~\cite{IBMcourse,Hughes,Ferris,Wong} are available for introducing quantum computing to students with minimal assumptions regarding background. We hope the present work demonstrates how a familiarity with undergraduate-level quantum mechanics can be leveraged to acquire expertise in using this exciting, emerging technology.

 \begin{acknowledgments}
The authors acknowledge the use of IBM Quantum services for this work. The views expressed are those of the authors, and do not reflect the official policy or position of IBM or the IBM Quantum team. The authors acknowledge the access to advanced services provided by the IBM Quantum Researchers Program.\cite{IBMQR} Additionally, the authors are grateful for insightful comments and valuable suggestions from the anonymous reviewers. The authors are particularly indebted to Raina Olsen for extensive, thoughtful feedback on previous versions of this manuscript. 

\end{acknowledgments}

\end{document}